\documentclass[11pt,intlimits,reqno]{amsart}

\usepackage{amsmath}

\usepackage{enumerate}

\usepackage{amssymb}

\usepackage{amsthm}

\usepackage{bbm}

\usepackage{graphicx}

\usepackage{mathrsfs}

\usepackage[polish,english]{babel}
\usepackage{verbatim}
\usepackage[utf8]{inputenc}
\usepackage{bbm}
\DeclareMathOperator\arcsinh{arcsinh}

\theoremstyle{definition}

\theoremstyle{remark}

\numberwithin{equation}{section}

\begin{document}

\title{An integrable (classical and quantum) four-wave mixing Hamiltonian system}

 \begin{abstract}
A four-wave mixing Hamiltonian system on the classical as well as on the quantum level is investigated. In the classical case, if one assumes the frequency resonance condition of the form $\omega_0 -\omega_1 +\omega_2 -\omega_3=0$, this Hamiltonian system is integrated in quadratures and the explicit formulas of solutions are presented. Under the same condition the spectral decomposition of quantum Hamiltonian is found and thus, the Heisenberg equation for this system is solved. Some applications of the obtained results in non-linear optics are disscused.
 \end{abstract}

\maketitle

\begin{center}
A. Odzijewicz, E. Wawreniuk\\
Institute of Mathematics\\
University in Bia\l{}ystok\\
Cio\l{}kowskiego 1M, 15-245 Bia\l{}ystok, Poland\\
aodzijew@uwb.edu.pl , ewawreniuk@math.uwb.edu.pl
\end{center}

\tableofcontents

\section{Introduction}

Interaction of four waves (four modes in quantum case) through a non-linear medium arises in many branches of physics including mechanics, optics, solid body physics and quantum information \cite{boyd,imam,jing,milburn,skryp2}. For example, in optics the four-wave mixing process describes the annihilation of two photons of frequencies $\omega_0$ and $\omega_2$ with the simultaneous creation of two photons of frequencies $\omega_1$ and $\omega_3$ while keeping energy and momentum  conserved. We mention also Raman and Brillouin processes, where in the second case photons interact through crystal lattice in a non-linear way with phonons and magnons \cite{jing, perina}. Except of a few integrated cases \cite{jurco, skryp, skryp2}, for the description of these phenomena one usually uses numerical \cite{Fle} or approximate methods \cite {bab,jurco, skryp, skryp2} which combine classical and quantum approaches.

This paper is the continuation of series of papers \cite{GOHS, OTH1,OTH2,OH4, KS, 3mod}, where the authors have studied the quantum and classical Hamiltonian systems applying the theory of orthogonal polynomials as well as the method of quantum and classical reduction. The model considered here, such as the models investigated in the previous papers, is chosen having in mind its usefulness for modeling non-linear phenomena in quantum and classical physics, particularly in optics.

We investigate the four-wave mixing system given in the classical case by the Hamiltonian \eqref{H} and in the quantum one by the Hamiltonian \eqref{qH1}. Assuming frequency resonance condition \eqref{conresonant} we integrate this system in both cases. Namely, in Section \ref{sec1} we present explicit formulas (\ref{solution1a}-\ref{solution2c}) for classical solutions. While in Section \ref{sec2}, applying the theory of dual Hahn polynomials, we obtain the spectral decomposition \eqref{Hdecomposition} of the quantum Hamiltonian  and thus, solve the Heisenberg equation, see \eqref{expqh1} and \eqref{evHeis}.

In Section \ref{sec3}, using standard coherent states, the correspondence between classical and quantum cases is shown in the limit $\hbar \to 0$. Next, in Section \ref{sec4} the reduced coherent states, see \eqref{redstate}, are described and the resolution of identity for these states is found (\ref{resp}- \ref{measure}).

In Section \ref{sec5} we show that the Hamiltonian \eqref{qH1}, after  rewriting it in terms of quantum angular momentum components, assumes the forms \eqref{qHM1} and \eqref{qHM2}. This allows us to interpret it as a Hamiltonian describing the interaction of two photons  with the quantum system composed of $N$ two-level atoms in the case \eqref{qHM1} and the interaction of two quantum angular momenta in the case \eqref{qHM2}.

\section{Classical four-wave mixing system}\label{sec1}

In this section we will  study the classical four-wave mixing Hamiltonian system on the phase space $\Omega_4 := \{ (z_0, z_1, z_2, z_3) \in \mathbb{C}^4 : |z_k|>0 \mbox{ for } k=0,1,2,3\}$ equipped with the Poisson bracket
\begin{equation}\label{pb}
\{ f, g \} :=i \sum_{k=0}^3 \left(\frac{\partial f}{\partial \bar z_k} \frac{\partial g}{\partial  z_k}-\frac{\partial f}{\partial  z_k}\frac{\partial g}{\partial \bar z_k}\right)
\end{equation}
of $f,g \in C^\infty (\Omega_4)$, taking
\begin{multline}\label{H}
H = \omega_0 |z_0|^2 +\omega_1 |z_1|^2+ \omega_2 |z_2|^2 +\omega_3 |z_3|^3 + \\
g( |z_0|^2|z_3|^2 + |z_1|^2|z_2|^2 + z_0 \bar z_1 z_2 \bar z_3 + \bar z_0 z_1 \bar z_2 z_3 )
\end{multline}
as its Hamilton function, where frequencies $\omega_0, \omega_1, \omega_2, \omega_3$ and coupling constant $g$ are real numbers. Note here that the Poisson bracket  \eqref{pb} is defined by the symplectic form 
\begin{equation}\label{formc4}
\omega_4 = i \sum_{k=0}^3 dz_k \wedge d \bar z_k .
\end{equation}

In order to  integrate this system we apply the reduction procedure presented in \cite{KS}, where more general case was considered on the classical as well as on the quantum level.  Following \cite{KS}, we define new canonical variables 
\begin{equation}\label{clasvar}
\begin{array}{rl}
I_0 &:= |z_0|^2, \\
I_1 &:= |z_0|^2 + |z_1|^2, \\
I_2 &:= |z_2|^2 + |z_3|^2,\\
I_3 &:= |z_0|^2 - |z_2|^2, 
\end{array} \mbox{ and }
\begin{array}{rl}
\psi_0 &:= \varphi_0-\varphi_1 +\varphi_2 -\varphi_3,\\
\psi_1 &:= \varphi_1, \\
\psi_2 &:= \varphi_3, \\
\psi_3 &:= \varphi_3 - \varphi_2, 
\end{array}
\end{equation}
where $z_k = |z_k|e^{i\varphi_k}$ for $k=0,1,2,3$, i.e. the Poisson bracket \eqref{pb} written in this variables is given by 
\begin{equation}\label{pb3}
\{ f, g \} := \sum_{k=0}^3 \left(\frac{\partial f}{\partial  I_k} \frac{\partial g}{\partial  \psi_k}-\frac{\partial f}{\partial  \psi_k}\frac{\partial g}{\partial  \psi_k}\right) . 
\end{equation}
Note that according to definition of $\Omega_4$ we have $|z_k|>0$, so, variables $I_0, I_1, I_2$ and $I_3$ must satisfy the  inequalities
\begin{equation}\label{onis}
I_0>0, \quad I_0< I_1, \quad I_0>I_3, \quad I_0< I_2+I_3.
\end{equation}
Since $0\leq \varphi_k < 2\pi $, from \eqref{clasvar} it follows that 
\begin{equation}
-4\pi \leq \psi_0 < 4\pi , \quad 0\leq \psi_1 < 2\pi, \quad 0\leq \psi_2 <2\pi , \quad -2\pi \leq \psi_3 < 2\pi . 
\end{equation}

The Hamiltonian \eqref{H} written in term of coordinates $(I_0, I_1, I_2, I_3, \psi_0, \psi_1, \psi_2, \psi_3)$ assumes the form
\begin{multline}\label{H2}
H= (\omega_0-\omega_1 +\omega_2 -\omega_3) I_0 +\omega_1I_1 +\omega_3 I_2 + (\omega_3 - \omega_2)I_3+\\
 g[I_0(I_2+I_3 -I_0) + (I_1-I_0)(I_0-I_3)+
2\sqrt{\mathcal{G}_0(I_0, I_1, I_2, I_3)}\cos \psi_0],
\end{multline}
where
\begin{equation}\label{gfunction}
\mathcal{G}_0(I_0, \vec{I}) := I_0(I_1-I_0)(I_0-I_3)(I_2+I_3-I_0)
\end{equation}
and $\vec{I}=(I_1, I_2, I_3)^T\in \mathbb{R}^3$.
Using the Poisson bracket \eqref{pb3} one finds that $I_1, I_2, I_3$ are integrals of motion $\{H, I_k \} =0$ in involution $\{I_k, I_l \}= 0$, $k,l=1,2,3$, for the above Hamiltonian. The Hamiltonian flows generated by them are 
\begin{align}
\nonumber
\sigma_{I_1} (t) (z_0, z_1, z_2, z_3) = & (e^{it}z_0, e^{it}z_1, z_2, z_3 ), \\
\label{iflows}
\sigma_{I_2} (t) (z_0, z_1, z_2, z_3) = & (z_0, z_1, e^{it}z_2, e^{it}z_3 ),\\
\nonumber
\sigma_{I_3} (t) (z_0, z_1, z_2, z_3) = & (e^{it}z_0, z_1, e^{-it}z_2, z_3 ).
\end{align}
So, they define the action of three-dimensional torus $\mathbb{T}^3= \mathbb{S}^1 \times \mathbb{S}^1 \times \mathbb{S}^1$ on the phase space $\Omega_4$. For this action the map $\vec{I}:\Omega_4\to \mathbb{R}^3$ defined by 
\begin{equation}\label{mommap}
\vec{I}(z_0,z_1, z_2, z_3) := \left(\begin{array}{c}
I_1 (z_0, z_1, z_2, z_3)\\
I_2(z_0, z_1, z_2, z_3)\\
I_3(z_0, z_1, z_2, z_3)
\end{array}\right)
\end{equation}
is the momentum map  if one identifies $\mathbb{R}^3$ with the space dual to Lie algebra of $\mathbb{T}^3$.

The coordinates $I_0, I_1, I_2, I_3$, as well as $\psi_0$, are invariants of the action \eqref{iflows} of $\mathbb{T}^3$, while the phase coordinates $(\psi_1, \psi_2, \psi_3)$  are transformed by 
\begin{align}
\nonumber
\sigma_{I_1} (t) (\psi_1, \psi_2, \psi_3) = (\psi_1 +t, \psi_2, \psi_3), \\
\label{flownapsi}
 \sigma_{I_2} (t) (\psi_1, \psi_2, \psi_3) = (\psi_1 , \psi_2+t, \psi_3),\\
\nonumber
\sigma_{I_3} (t) (\psi_1, \psi_2, \psi_3) = (\psi_1 , \psi_2, \psi_3+t).
\end{align}
Hence, the quotient $\vec{I}^{-1}(\vec{b})/ \mathbb{T}^3$ of the level set 
$$\vec{I}^{-1}(\vec{b}) := \{ (z_0, z_1, z_2, z_3)\in \Omega_4 : \vec{I}(z_0, z_1, z_2, z_3) = \vec{b}\}$$
 by $\mathbb{T}^3$  is a two-dimensional manifold diffeomorphic $\vec{I}^{-1}(\vec{b})/ \mathbb{T}^3\cong ]a,b[\times \mathbb{S}^1$ to the cylinder $]a,b[\times \mathbb{S}^1$, where $a:= \max \{0, b_3\}$, $b:= \min \{ b_1, b_2+b_3 \}$ and $\vec{b}= (b_1, b_2, b_3)^T \in \vec{I}(\Omega_4)$.

After reduction to $]a,b[\times \mathbb{S}^1\cong \vec{I}^{-1}(\vec{b})/\mathbb{T}^3$ the Poisson bracket \eqref{pb3} of $f,g \in C^\infty (]a,b[\times \mathbb{S}^1)$ in the coordinates $(I_0, e^{i\psi_0})\in ]a,b[\times \mathbb{S}^1$ assumes the canonical form 
\begin{equation}\label{pb2}
\{f, g\}_{red} = \frac{\partial f}{\partial I_0}\frac{\partial g}{\partial \psi_0}-\frac{\partial g}{\partial I_0}\frac{\partial f}{\partial \psi_0} 
\end{equation}
and the Hamiltonian \eqref{H2} is given by 
\begin{multline}\label{H3}
H_{red}= (\omega_0-\omega_1 +\omega_2 -\omega_3) I_0 +\omega_1b_1 +\omega_3 b_2 + (\omega_3 - \omega_2)b_3+\\
 g[I_0(b_2+b_3 -I_0) + (b_1-I_0)(b_0-c_3)+
2\sqrt{\mathcal{G}_0(I_0, \vec{b})}\cos \psi_0].
\end{multline}
For the definition of  $\mathcal{G}_0(I_0, \vec{b})$ see \eqref{gfunction}.
Using \eqref{pb2} and \eqref{H3} we immediately obtain the Hamilton equations 
\begin{align}\label{di0}
\frac{d}{dt} I_0 &=-\frac{\partial  H_{red}}{\partial \psi_0} = 2g \sqrt{\mathcal{G}_0(I_0, \vec{b})}\sin \psi_0,\\
\label{dpsi0}
\frac{d}{dt} \psi_0 &= \frac{\partial H_{red}}{\partial I_0}= -4gI_0 + (\omega_0-\omega_1 +\omega_2 -\omega_3)+g(b_1+b_2+2b_3) +\\
\nonumber
  & g \frac{ \mathcal{G}_0'(I_0, \vec{b})}{\sqrt{\mathcal{G}_0(I_0, \vec{b})}}\cos \psi_0 
\end{align}
on the reduced phase space $]a,b[\times \mathbb{S}^1$.
Combining $H_{red}( I_0(t), \psi_0 (t)) = E = const$ with (\ref{di0}-\ref{dpsi0}) we find 
\begin{multline}\label{diffi0}
\left(\frac{d}{dt} I_0\right)^2 = 
 4g^2 \mathcal{G}_0(I_0, \vec{b})-
\Big(E- (\omega_0-\omega_1 +\omega_2 -\omega_3) I_0 -\omega_1b_1 -\omega_3 b_2 - (\omega_3 - \omega_2)b_3-\\
g[I_0(b_2+b_3 -I_0) +(b_1-I_0)(I_0-b_3)]\Big)^2, 
\end{multline}
\begin{multline}\label{diffpsi}
e^{i\psi_0} = \frac{1}{2g\mathcal{G}_0(I_0, \vec{b})}\Big(E - (\omega_0-\omega_1+\omega_2 -\omega_3)I_0 - \omega_1 b_1 - \omega_3b_2 - (\omega_3-\omega_2)b_3-\\
g(I_0(b_2+b_3-I_0) + (b_1-I_0)(I_0-b_3)) + i \frac{d}{dt}I_0\Big).
\end{multline}
Since on the right hand side of the equation \eqref{diffi0} one has a polynomial of the variable $I_0$ of the degree not bigger than four, the solution $I_0(t)$ of \eqref{diffi0} is expressed by the inverse of an elliptic function \cite{ryzhik}. Therefore, knowing $I_0(t)$ one finds $e^{i\psi_0(t)}$ from equation \eqref{diffpsi}.

The phase functions $ \psi_1 (t), \psi_2 (t), \psi_3 (t)$ one obtains integrating the equations
\begin{align}
\label{dpsi1}
\frac{d}{dt} \psi_1 (t) &=\frac{\partial H}{\partial I_1}(I_0(t), \psi_0(t))= g(I_0(t) -b_3) +\omega_1 + \\
\nonumber 
 & g\frac{-I_0(t)^3 +(2b_3+b_2)I_0(t)^2-(b_2+b_3)b_3I_0(t)}{\sqrt{\mathcal{G}_0(I_0(t), \vec{b})}}\cos \psi_0(t),\\
\frac{d}{dt} \psi_2(t) &=\frac{\partial H}{\partial I_2}(I_0(t), \psi_0(t))= gI_0(t) + \omega_3 + \\
\nonumber
  & g\frac{-I_0(t)^3 +(b_1+b_3)I_0(t)^2-b_1b_3I_0(t)}{\sqrt{\mathcal{G}_0(I_0(t), \vec{b})}}\cos \psi_0(t),\\
\label{dpsi3}
\frac{d}{dt} \psi_3 (t)&=\frac{\partial H}{\partial I_3}(I_0(t), \psi_0(t))= 2gI_0(t)-gb_1 + \omega_3-\omega_2 +\\
\nonumber
 & g\frac{-2I_0(t)^3 +(2b_1+b_2+2b_3)I_0(t)^2-(b_1b_2+2b_1b_3)I_0(t)}{\sqrt{\mathcal{G}_0(I_0(t), \vec{b})}}\cos \psi_0(t),\\
\end{align}
whose right hand sides depend on known functions $I_0(t)$ and $e^{i\psi_0(t)}$ only. We recall here that $I_k(t) = b_k$ for $k=1,2,3$.

Now, assuming the frequency resonance condition 
\begin{equation}\label{conresonant}
\omega_0 - \omega_1 +\omega_2 - \omega_3 =0,
\end{equation}
relevant from a physical point of view, we will find the explicit forms of $I_0(t)$ and $e^{i\psi_0(t)}$. Namely, under this condition the right hand side of \eqref{diffi0} reduces to a polynomial of degree non greater than two, so, we have
\begin{equation}\label{inti0}
\left(\frac{d}{dt}I_0\right)^2 = pI_0^2 +qI_0 +r, 
\end{equation}
where  the constants $p,q$ and $r$ are defined as follows 
\begin{align}
\nonumber
p:= & -g^2(b_1-b_2)^2 - 4g(E-\omega_1b_1 - \omega_3b_2 -(\omega_3-\omega_2)b_3),\\
q:= & 2g^2b_1b_3 (b_1-b_2)+2g(b_1+b_2+2b_3)(E-\omega_1b_1 - \omega_3b_2 -(\omega_3-\omega_2)b_3),\\
\nonumber
r := & -(E-\omega_1b_1 - \omega_3b_2 -(\omega_3-\omega_2)b_3+gb_1b_3)^2.
\end{align}
Thus, if \eqref{conresonant} is satisfied, the solutions of equations \eqref{inti0} and \eqref{diffpsi} are given by:\\
a)  if $\Delta = q^2 - 4pr >0$ and $ p<0$, then
\begin{align}\label{solution1a}
I_0(t) = & \frac{\sqrt{\Delta }}{2p} \sin \left(-\sqrt{-p}(t-t_0)+C\right)- \frac{q}{2p},\\
\nonumber
e^{i\psi_0(t)} = &\frac{1}{2g\mathcal{G}_0(I_0(t), \vec{b})}\Big(E -  \omega_1 b_1 - \omega_3b_2 - (\omega_3-\omega_2)b_3-\\
\nonumber
 &g[I_0(t)(b_2+b_3-I_0(t)) + (b_1-I_0(t))(I_0(t)-b_3)] -\\
 & i \frac{\sqrt{-p\Delta }}{2p} \cos \left(-\sqrt{-p}(t-t_0)+C\right)\Big),
\end{align}
b) if $\Delta =0, p>0$, then
\begin{align}
I_0(t) = &C\exp (\sqrt{p}(t-t_0)) -\frac{q}{2p}, \\
\nonumber
e^{i\psi_0(t)} = &\frac{1}{2g\mathcal{G}_0(I_0(t), \vec{b})}\Big(E -  \omega_1 b_1 - \omega_3b_2 - (\omega_3-\omega_2)b_3-\\
 &g[I_0(t)(b_2+b_3-I_0(t)) + (b_1-I_0(t))(I_0(t)-b_3)] + i C\sqrt{p}\exp (\sqrt{p}(t-t_0))\Big),
\end{align}
c) if $\Delta <0, p>0$, then 
\begin{align}
I_0(t) = &\frac{\sqrt{-\Delta }}{2p} \sinh \left(\sqrt{p}(t-t_0)+C\right)- \frac{q}{2p},\\ 
\nonumber
e^{i\psi_0(t)} = &\frac{1}{2g\mathcal{G}_0(I_0(t), \vec{b})}\Big(E -  \omega_1 b_1 - \omega_3b_2 - (\omega_3-\omega_2)b_3-\\
\nonumber
 &g[I_0(t)(b_2+b_3-I_0(t)) + (b_1-I_0(t))(I_0(t)-b_3)] + \\
\label{solution2c}
 &i \frac{\sqrt{-p\Delta }}{2p} \cosh \left(\sqrt{p}(t-t_0)+C\right)\Big).
\end{align}
For the above cases, the constant $C$ depends on the initial condition $(t_0, I_0(t_0))$ in the following way
\begin{displaymath}
C:= \begin{cases}
\arcsin \left(\frac{2pI_0(t_0) +q}{\sqrt{-\Delta}}\right)\mbox{ if } \Delta>0, p<0,\\
I_0(t_0) + \frac{q}{2p} \mbox{ if } \Delta=0, p>0,\\
\arcsinh \left(\frac{2pI_0(t_0) +q}{\sqrt{\Delta}}\right)\mbox{ if } \Delta<0, p>0 .
\end{cases}
\end{displaymath}

There exists another way to find the trajectories of the obtained above Hamiltonian evolutions $\mathbb{R} \ni t \mapsto (I_0(t), \psi_0 (t))\in ]a,b[\times \mathbb{S}^1$, which does not demand computing any integrals. It is based on the  realization of reduced phase space $\vec{I}^{-1}(\vec{b})/ \mathbb{T}^3\cong ]a, b[ \times \mathbb{S}^1$ as a circularly symmetric surface in $\mathbb{R}^3$ called  further Kummer shape \cite{holm, KS}. For this reason, let us define a new complex variable 
\begin{equation}\label{z}
z= x+iy:=  z_0 \bar z_1 z_2 \bar z_3 . 
\end{equation}

Using the variables $(x,y ,I_0)^T \in \mathbb{R}^3$ as the coordinates on $\mathbb{R}^3$, we define the  map 
\begin{equation}\label{onKummer}
\Phi_{\vec{c}} (I_0, \psi_0) := \left(\begin{array}{c}
 \sqrt{\mathcal{G}_0(I_0, \vec{b})}\cos \psi_0\\
 \sqrt{\mathcal{G}_0(I_0, \vec{b})}\sin \psi_0\\
I_0
\end{array}\right)=\left(\begin{array}{c}
x\\
y\\
I_0
\end{array}\right) 
\end{equation}
of  $\vec{I}^{-1}(\vec{b})/ \mathbb{T}^3\cong ]a,b[\times \mathbb{S}^1$ into $\mathbb{R}^3$. We also define the Nambu-Poisson bracket 
\begin{equation}\label{nambu}
\{f, g \}_{\mathcal{C}} := \det [\nabla \mathcal{C}, \nabla f , \nabla g ],
\end{equation}
of $f,g \in C^\infty (\mathbb{R}^3)$, where $\nabla f = \left(\frac{\partial f}{\partial x},\frac{\partial f}{\partial y}, \frac{\partial f}{\partial I_0}\right)^T$, and 
\begin{equation}
\mathcal{C}(x,y, I_0) := \frac{1}{2}(\mathcal{G}_0(I_0, \vec{b})-(x^2+y^2)) .
\end{equation}

Obviously, $\mathcal{C}$ is a Casimir function for the Poisson-Nambu bracket \eqref{nambu}, i.e. $\{\mathcal{C} , f\}_{\mathcal{C}} = 0$ for every $f\in C^\infty (\mathbb{R}^3)$. 
Therefore, the circularly symmetric surfaces $\mathcal{C}^{-1}(\lambda)\backslash \{ I_0 \in \mathbb{R} : \frac{d \mathcal{G}_0}{dI_0} (I_0, \vec{b}) =0 \mbox{ and } \mathcal{G}_0(I_0,\vec{b}) +\lambda = 0\}$, where $\lambda \in \mathbb{R}$, are symplectic leaves of the Poisson manifold $(C^\infty (\mathbb{R}^3), \{ \cdot, \cdot \}_{\mathcal{C}})$.

One easily sees that $\Phi_{\vec{c}}:]a,b[\times \mathbb{S}^1 \to \mathbb{R}^3$  is a Poisson map, i.e. 
\begin{equation}
\{f \circ \Phi_{\vec{c}}, g\circ \Phi_{\vec{c}}\}_{red} = \{f,g\}_{\mathcal{C}} \circ \Phi_{\vec{c}}  
\end{equation}
and thus, $\Phi_{\vec{c}}:]a,b[\times \mathbb{S}^1\to \mathcal{C}^{-1}(0)\backslash \{(0,0,a)^T, (0,0,b)^T\}$ is a symplectic diffeomorphism of the reduced symplectic manifold $]a,b[ \times \mathbb{S}^1 $ on the symplectic leaf $ \mathcal{C}^{-1}(0)\backslash \{(0,0,a)^T, (0,0,b)^T\}$.

Note that the  functions $I_0, I_1, I_2, I_3, x, y \in C^\infty (\Omega_4)$ generate a Poisson subalgebra $C^\infty_{\mathcal{G}_0}(\Omega_4)$ of  $(C^\infty (\Omega_4), \{\cdot, \cdot \})$ for which the integrals of motion $I_1, I_2$ and $I_3$ are Casimir functions, i.e.  the following relations
\begin{align}
\label{na1}
\{I_0, x\} = &-y, \\
\{I_0, y\} = &x,\\
\{ x,  y \} = &\frac{1}{2}   \frac{\partial \mathcal{G}_0}{\partial I_0}(I_0, I_1, I_2, I_3),\\
\label{na3}
\{I_k, x\}= &\{I_k, y \} = \{I_k, I_0\} = 0, \mbox{ for } k=1,2,3   
\end{align} 
are satisfied.
Reducing above relations to $\vec{I}^{-1} (\vec{b})$, i.e. substituting the constant $\vec{b}$ instead of the function $\vec{I}$ into (\ref{na1}-\ref{na3}), one finds that the coordinate functions $x,y,I_0$ satisfy  the relations
\begin{align}
\{I_0, x\}_{\mathcal{C}} = & -y, \quad \{I_0, y \}_{\mathcal{C}} = x , \\
\{x,y \}_{\mathcal{C}} = &\frac{1}{2} \frac{\partial \mathcal{G}_0}{\partial I_0}(I_0, \vec{c}), \\
\{\mathcal{C},x\}_{\mathcal{C}} =& \{\mathcal{C},y\}_{\mathcal{C}}=\{\mathcal{C},I_0\}_{\mathcal{C}} =0
\end{align}
with respect to the Nambu-Poisson bracket. The above means that the Poisson subalgebra  of $(C^\infty (\Omega_4), \{ \cdot , \cdot \})$ generated by functions $x,y, I_0 \in C^\infty (\Omega_4)$ is isomorphic  to the Poisson algebra $(C^\infty (\mathbb{R}^3) , \{ \cdot, \cdot \}_{\mathcal{C}})$ of smooth functions on $\mathbb{R}^3$ with $\{\cdot , \cdot \}_{\mathcal{C}}$ as a Poisson bracket.

From the above facts, we see that the reduced Hamiltonian  \eqref{H3} written in terms of $x,y, I_0$ can be treated as a Hamiltonian
\begin{multline}\label{H4}
H_{red}=  (\omega_0-\omega_1 +\omega_2 -\omega_3) I_0 +\omega_1b_1 +\omega_3b_2 + (\omega_3 - \omega_2)b_3+\\
 g[I_0(b_2+b_3 -I_0) + (b_1-I_0)(I_0-b_3)+2x]
\end{multline}
from $(C^\infty(\mathbb{R}^3), \{\cdot , \cdot \}_{\mathcal{C}})$.
Hence, a trajectory  of the evolution $\mathbb{R}\ni t \mapsto (x(t), y(t), I_0(t))$ defined by $H_{red}$ is the intersection of the Kummer shape $\mathcal{C}^{-1}(0)\backslash \{(0,0,a)^T, (0,0,b)^T\}$  with a level set $H_{red}^{-1} (E)$ of the Hamiltonian \eqref{H4}. For the more detailed investigation of evolution trajectories for a four-wave mixing system in terms of Kummer shape we address to \cite{OTT}. See also \cite{3mod}, where the three-wave case is considered. A hierarchy of integrable Hamiltonians describing $n$-wave non-linear interaction is presented in \cite{GO}.

\section{Quantum four-wave mixing system}\label{sec2}

As the quantum counterpart of the classical Hamiltonian \eqref{H} we take
\begin{multline}\label{qH1}
\hat{H} = \omega_0 a_0^*a_0 +\omega_1 a_1^*a_1 +\omega_2 a_2^*a_2 +\omega_3 a_3^*a_3 + \\ g[  a_0^*a_0 a_3a_3^*  +  a_1^*a_1 a_2a_2^*+  a_0a_1^* a_2 a_3^* + a_0^* a_1 a_2^* a_3 ]=
\omega_0 a_0^*a_0 +\omega_1 a_1^*a_1 +\omega_2 a_2^*a_2 +\omega_3 a_3^*a_3 + \\ g[  a_0^*a_0 (a_3^*a_3 +\hbar) +  a_1^*a_1 (a_2^*a_2 +\hbar)
+  a_0a_1^* a_2 a_3^* + a_0^* a_1 a_2^* a_3 ].
\end{multline}
Having in mind physical applications, we will keep the Planck constant $\hbar$  also in the  further expressions. Hence, the anihilation $a_k$ and creation $a_k^*$, $k=0,1,2,3$, operators written  in the Fock basis $|n_0, n_1, n_2, n_3 \rangle$, $n_0, n_1, n_2, n_3 \in \mathbb{N}\cup \{0\}$, of the corresponding Hilbert space $\mathcal{H}$ are given by 
\begin{align}
a_k |n_0, n_1, n_2, n_3 \rangle = &\sqrt{\hbar n_k } |n_0, \ldots, n_k-1, \ldots, n_3\rangle ,\\
a_k^* |n_0, n_1, n_2, n_3 \rangle = &\sqrt{\hbar (n_k+1) } |n_0, \ldots, n_k+1, \ldots, n_3\rangle 
\end{align}
and $a_k |0,0,0,0\rangle = 0$, where $k=0,1,2,3$, so, they obey the standard commutation relations 
\begin{equation}\label{eqs1}
[a_k, a_l^* ] = \hbar\delta_{kl} , \quad [a_k, a_l ] = [a_k^*, a_l^*] =0 \quad \mbox{ for } k,l = 0,1,2,3 . 
\end{equation}

Now, by analogy to the classical case, see \eqref{clasvar} and \eqref{z}, we introduce new quantum variables:
\begin{align}\label{qvariables1}
A_0 &:= a_0^* a_0,\\
\label{qvariables2}
A_1 &:= a_0^*a_0 + a_1^*a_1,\\
\label{qvariables3}
A_2 &:= a_2^*a_2 + a_3^*a_3,\\
\label{qvariables4}
A_3 &:= a_0^*a_0- a_2^*a_2,\\
A &:=  a_0a_1^* a_2 a_3^* , \\
\label{qvariables5}
A^* &:=  a_0^* a_1 a_2^* a_3 , 
\end{align}
which satisfy the following commutation relations
\begin{align}\label{coma0a}
[A_0, A] &= -\hbar A , \quad [A_0, A^*] = \hbar A^* , \\
\label{com00}
[A_k, A_0] &= [A_k, A]= [A_k, A^*] = 0 \mbox{ for } k=1,2,3,\\
\label{relcom}
[A, A^* ] = &\mathcal{G}_\hbar (A_0 +\hbar , A_1, A_2, A_3) - \mathcal{G}_\hbar (A_0, A_1, A_2, A_3).
\end{align} 
Let us mention that the relation \eqref{relcom} follows from 
\begin{align}
A^* A =& \mathcal{G}_\hbar (A_0, A_1, A_2, A_3),\\
AA^* = &\mathcal{G}_\hbar (A_0+ \hbar, A_1, A_2, A_3),
\end{align}
where 
\begin{equation}
\mathcal{G}_\hbar (A_0, A_1, A_2, A_3) := A_0 (A_1 - A_0 +\hbar)(A_0 - c_3)(A_2+A_3 - A_0 +\hbar).
\end{equation}

The operator algebra $\mathcal{A}_{\mathcal{G}_\hbar}$ generated by $A_0, A_1, A_2, A_3, A$ and $A^*$, which  satisfy the relations (\ref{coma0a}-\ref{relcom}), could be considered as a quantum counterpart to the Poisson algebra $C^\infty_{\mathcal{G}_0}(\Omega_4)$ defined by (\ref{na1}-\ref{na3}). In the next section we describe the correspondence between these algebras. 
Note also, that $A_0, A_1, A_2, A_3$ are diagonal in the Fock basis, i.e.
\begin{equation}\label{akdiag}
A_k |n_0, n_1, n_2, n_3 \rangle = \hbar c_k |n_0, n_1, n_2, n_3 \rangle , 
\end{equation}
where according to (\ref{qvariables1}-\ref{qvariables4}) their eigenvalues measured in $\hbar$-units are given by 
\begin{equation}\label{cn}
c_0 = n_0, \quad c_1 = n_0+n_1, \quad c_2 = n_2+n_3, \quad c_3 = n_0-n_2
\end{equation}
and thus, $c_0,c_1, c_2, c_3 \in \mathbb{Z}$ satisfy inequalities
\begin{equation}\label{216}
c_0=n_0 \geq 0, \quad c_1-c_0= n_1  \geq 0,\quad  c_0-c_3 =n_2 \geq 0, \quad c_2+c_3 - c_0 = n_3\geq 0,
\end{equation}
which are equivalent to the positivity conditions 
\begin{equation}\label{positivitycon}
A_0 \geq 0, \quad A_1-A_0\geq 0,\quad  A_0-A_3 \geq 0, \quad A_2+A_3 - A_0 \geq 0
\end{equation}
for the corresponding operators. Let us note that these conditions are the same as the ones in \eqref{onis} for the classical counterparts $I_0, I_1, I_2$ and $I_3$ of these operators.

Rewriting Hamiltonian \eqref{qH1} in terms of $A_0, A_1, A_2, A_3, A$ and $A^*$ one obtains
\begin{multline}\label{qH2}
\hat{H}=  (\omega_0 -\omega_1 +\omega_2 -\omega_3) A_0 + \omega_1 A_1 +\omega_3 A_2 + (\omega_3 - \omega_2)A_3+\\
 g(A_0 (A_2+A_3-A_0 +\hbar) + (A_1-A_0)(A_0-A_3+\hbar)+  A +   A^*).
\end{multline}
Hence, from (\ref{coma0a}-\ref{com00}) and \eqref{qH2} it is easy to see that $[A_k, \hat{H}]=0$ for $k=1,2,3$, so, the operators $A_1, A_2, A_3$ are quantum integrals of motion for the system described by the Hamiltonian $\hat{H}\in \mathcal{A}_{\mathcal{G}_{\hbar}}$ defined in \eqref{qH2}. Therefore one can reduce this quantum system to their common  eigensubspaces $\mathcal{H}_{\vec{c}} \subset \mathcal{H}$ parametrized by the corresponding eigenvalues $ \vec{c} := (c_1, c_2, c_3)^T\in C_3$, where the cone $C_3 \subset \mathbb{Z}^3$ is defined by  
\begin{equation}\label{219}
C_3 := \{ \vec{c} \in \mathbb{Z}^3 : c_1 \geq 0, c_2 \geq 0, c_1 - c_3 \geq 0 \mbox{ and } c_2 + c_3 \geq 0 \}. 
\end{equation}
Let us note here that inequalities in \eqref{219} follow from \eqref{positivitycon}. From \eqref{216} we conclude that the Fock vectors 
\begin{equation}\label{basisnew}
| n, c_1 -n, -c_3 +n, c_2 + c_3 -n \rangle ,
\end{equation}
where 
\begin{equation}\label{nvariety}
\max \{ 0, c_3 \} \leq n \leq \min \{ c_1, c_2 +c_3 \}, 
\end{equation}
form an orthogonal basis of the eigensubspace $\mathcal{H}_{\vec{c}}$. Thus, one obtains the formula $\dim \mathcal{H}_{\vec{c}} = N+1$ on the dimension  of $\mathcal{H}_{\vec{c}}$, where 
\begin{equation}
N= \min \{ c_1, c_2 +c_3 \}-\max \{ 0, c_3 \}. 
\end{equation}
Taking into account \eqref{nvariety}, we distinguish the four possible subcases:
\begin{align}
\label{cases1}
\mbox{(i) } & \max\{ 0, c_3 \} = 0 \mbox{ and } \min\{c_1, c_2+c_3 \} =c_1 , \\
\mbox{(ii) } & \max\{ 0, c_3 \} = 0 \mbox{ and } \min\{c_1, c_2+c_3 \} =c_2+c_3, \\
\mbox{(iii) } & \max\{ 0, c_3 \} = c_3\mbox{ and } \min\{c_1, c_2+c_3 \} =c_1 ,\\
\label{cases4}
\mbox{(iv) } & \max\{ 0, c_3 \} = c_3 \mbox{ and } \min\{c_1, c_2+c_3 \} =c_2+c_3 . 
\end{align}
Let us note here that these subcases are not disjoint, i.e. for some $\vec{c}\in C_3$ the subspace $\mathcal{H}_{\vec{c}}$ could belong to more than one subclasses.

In order to have the common description of these subcases, we numerate the basis \eqref{basisnew} of $\mathcal{H}_{\vec{c}}$ as follows
\begin{equation}\label{newbasis}
|n, N-n , \gamma +n, N +\delta -n \rangle =: |n\rangle , 
\end{equation}
where $n=0,1,\ldots, N$  and the other integer parameters $N,\gamma$ and $\delta$  linearly depend 
\begin{align}
\label{case1}
\mbox{(i) } &N=c_1, \gamma = -c_3, \delta = c_2 + c_3 - c_1, \\
\mbox{(ii) } & N=c_2+c_3, \gamma = -c_3, \delta = c_1 - c_2 - c_3, \\
\mbox{(iii) } &N=c_1- c_3 , \gamma = c_3, \delta = c_2 + c_3 - c_1, \\
\label{case4}
\mbox{(iv) } &N=c_2, \gamma = c_3, \delta = c_1 - c_2 - c_3
\end{align}
on  $\vec{c} = (c_1, c_2, c_3)^T \in C_3$.

Let $P_{\vec{c}}: \mathcal{H} \to \mathcal{H}_{\vec{c}}$ be the orthogonal projection of $\mathcal{H}$ on the Hilbert subspace $\mathcal{H}_{\vec{c}}$. Arbitrary element $\hat{F} \in \mathcal{A}_{\mathcal{G}_{\hbar}}$ of the algebra $\mathcal{A}_{\mathcal{G}_{\hbar}}$  commutes $\hat{F} P_{\vec{c}} = P_{\vec{c}} \hat{F}$ with $P_{\vec{c}}$. So, the  operators $A_0, A$ and $A^*$ split into operators
\begin{equation}\label{eq335}
\textbf{A}_{0\vec{c}} := \frac{1}{\hbar} P_{\vec{c}}A_0 P_{\vec{c}}, \quad \textbf{A}_{\vec{c}} := \frac{1}{\hbar^2} P_{\vec{c}}A P_{\vec{c}}, \quad \textbf{A}^*_{\vec{c}}:= \frac{1}{\hbar^2} P_{\vec{c}}A^* P_{\vec{c}}.
\end{equation}
According to the terminology assumed in the classical case, we will call the reduced quantum algebra, i.e. the algebra $\mathcal{A}_{\mathcal{G}_{\hbar}, \vec{c}}$ generated by $\textbf{A}_{0\vec{c}}, \textbf{A}_{\vec{c}}$ and $\textbf{A}^*_{\vec{c}}$, the quantum Kummer shape algebra. The operators defined in \eqref{eq335} act on the basis  \eqref{newbasis} by
\begin{equation}\label{a}
\textbf{A}_{\vec{c}} |n\rangle = \sqrt{n(N-n+1)(\gamma + n)(N-n + \delta +1)} |n-1\rangle , 
\end{equation}
\begin{equation}\label{astar}
\textbf{A}^*_{\vec{c}} |n\rangle = \sqrt{(n+1)(N-n)(\gamma + n+1) (N-n +\delta )} |n+1\rangle  
\end{equation}
and by
\begin{equation}\label{a0}
\textbf{A}_{0 \vec{c}} |n \rangle = n|n\rangle \mbox{ or } \textbf{A}_0 |n \rangle = (c_3+n)|n\rangle ,
\end{equation}
where the first equality in \eqref{a0} is taken for the subcases (i) and (ii), and the second one for the subcases (iii) and (iv). Assuming that $\textbf{A}_{0\vec{c}}$ is given by the first formula in \eqref{a0} we obtain the expression  
\begin{equation}\label{qH3}
\textbf{H}_{\vec{c}} = \frac{1}{\hbar} P_{\vec{c}}\hat{H} P_{\vec{c}}= (\omega_0 -\omega_1 +\omega_2 -\omega_3) \textbf{A}_{0\vec{c}} + g\hbar\textbf{H}_{0\vec{c}} + \lambda_{0\vec{c}},
\end{equation}
for the reduced Hamiltonian common for all these subcases, where 
\begin{multline}\label{qH4}
\textbf{H}_{0\vec{c}} :=  \textbf{A}_{0\vec{c}} ( N - \textbf{A}_{0\vec{c}} + \delta + 1) + 
( N - \textbf{A}_{0\vec{c}})( \gamma + \textbf{A}_{0\vec{c}} + 1 )  + \textbf{A}_{\vec{c}} + \textbf{A}^*_{\vec{c}}  , 
\end{multline}
and the constant $\lambda_{0\vec{c}}$ is equal to 
\begin{align}
\label{1case}
\mbox{(i) } \lambda_{0\vec{c}} = &\omega_1  c_1 + \omega_3 c_2 + (\omega_3 - \omega_2 )c_3,\\
\label{2case}
\mbox{(ii) } \lambda_{0\vec{c}} = & \omega_1  c_1 + \omega_3 c_2 + (\omega_3 - \omega_2 )c_3 + g\hbar(c_1 -c_2 -c_3)(1-c_3), \\
\label{3case}
\mbox{(iii) } \lambda_{0\vec{c}} = & \omega_1  c_1 + \omega_3 c_2 + (\omega_3 - \omega_2 )c_3 + g\hbar c_3 (c_2 + c_3 -c_1 +1), \\
\label{4case}
\mbox{(iv) } \lambda_{0\vec{c}} =  &\omega_1  c_1 + \omega_3 c_2 + (\omega_3 - \omega_2 )c_3 + g\hbar (c_1 - c_2) , 
\end{align}
respectively.
Let us stress that Hamiltonian \eqref{qH3} and the constant $g\hbar$ have dimension of the inverse of time. In a consequence, the operators $\textbf{A}_{0\vec{c}}, \textbf{A}_{\vec{c}}, \textbf{A}^*_{\vec{c}}$ and $\textbf{H}_{0\vec{c}}$ are dimensionless.

One easily sees that in the basis \eqref{newbasis} the operator $\textbf{H}_{0\vec{c}}$  has the three-diagonal form 
\begin{equation}
\textbf{H}_{0\vec{c}} |n \rangle = b_{n-1} |n-1\rangle + a_n |n\rangle + b_n |n+1\rangle , 
\end{equation}
where 
\begin{equation}
a_n = n(N-n+\delta +1)+ (N-n)(\gamma +n+1), 
\end{equation}
\begin{equation}
b_n = \sqrt{(n+1)(N-n)(\gamma+n+1)(N-n+\delta)}, 
\end{equation}
are the coefficients of the three-term recurrence 
\begin{equation}
\lambda R_n (\lambda ; \gamma, \delta , N) = b_{n-1} R_{n-1} (\lambda; \gamma , \delta N) + a_n R_n (\lambda; \gamma , \delta , N) + b_n R_{n+1}(\lambda; \gamma , \delta, N)
\end{equation}
of the dual Hahn polynomials
\begin{multline}\label{dHp}
R_n (\lambda; \gamma, \delta, N) :=\sqrt{{\gamma +n\choose n}{\delta +N-n\choose N-n}}\sum_{j=0}^\infty \left(\frac{(-n)_j  }{(\gamma +1)_j (-N)_j j!} \prod_{l=0}^{j-1} (-\lambda +l(l+\gamma +\delta +1))\right) 
\end{multline}
of the variable $\lambda$ \cite{chich, koek}. Dual Hahn polynomials are orthogonal 
\begin{equation}
\int_{\mathbb{R}} R_n (\lambda; \gamma , \delta , N ) R_m (\lambda ; \gamma , \delta, N) d\mu (\lambda) = \delta_{nm}
\end{equation}
with respect to the finite support measure 
\begin{equation}
d\mu (\lambda) = \sum_{k=0}^N \delta ( \lambda - \lambda_k) d\lambda ,
\end{equation}
where $\lambda_k$ are given by 
\begin{equation}
\lambda_k = k(k+\gamma +\delta +1) \mbox{ for } k=0,1, \ldots , N. 
\end{equation}
Therefore, using the theory of finite orthogonal polynomials, we find that 
\begin{equation}
\textbf{H}_{0\vec{c}} |\lambda_k \rangle = \lambda_k |\lambda_k \rangle ,
\end{equation}
where the eigenvector $|\lambda_k\rangle $ is given by  
\begin{equation}
|\lambda_k \rangle = \sum_{n=0}^N R_n (\lambda_k; \gamma, \delta, N) |n\rangle . 
\end{equation}
The values of dual Hahn polynomials \eqref{dHp} taken at $\lambda_k$ are the following 
\begin{multline}\label{dualH}
R_n (\lambda_k; \gamma, \delta, N) := \sqrt{{\gamma +n\choose n}{\delta +N-n\choose N-n}} \mbox{ }_3F_2\left(\left.\begin{array}{c}
-n, -k, k+\gamma+\delta+1 \\
\gamma +1 , -N 
\end{array}\right|  1\right)=\\
\sqrt{{\gamma +n\choose n}{\delta +N-n\choose N-n}}\sum_{j=0}^\infty \frac{(-n)_j (-k)_j (k+\gamma + \delta +1)_j }{(\gamma +1)_j (-N)_j} \frac{1}{j!}.
\end{multline}
From the orthogonality property 
\begin{equation}
\langle \lambda_k |\lambda_l \rangle = \delta_{kl}\langle \lambda_k | \lambda_k \rangle  
\end{equation}
of eigenvectors $|\lambda_k\rangle $ and from $ \langle n|m \rangle = \delta_{nm}$ one finds that the $(N+1)\times (N+1)$ matrix $\textbf{R}_{\vec{c}}= [R_{kn}^{\vec{c}}]$, defined by
\begin{equation}\label{rmatrix}
R_{nk}^{\vec{c}} := \langle \lambda_k | \lambda_k \rangle^{-\frac{1}{2}} R_n (\lambda_k; \gamma , \delta , N ), 
\end{equation}
satisfies $\textbf{R}_{\vec{c}}\textbf{R}_{\vec{c}}^T=\mathbbm{1}$.  This orthogonal matrix gives transition between the orthogonal bases $\{|n \rangle \}_{n=0}^N $ and $\{\langle \lambda_k |\lambda_k \rangle^{-\frac{1}{2}} |\lambda_k \rangle \}_{k=0}^N$.

For dual Hahn polynomials one has 
\begin{equation}\label{measure}
\langle \lambda_k |\lambda_k \rangle = \frac{(-1)^k k! (\delta+1)_k (k+\gamma +\delta +1)_{N+1}}{N! (-N)_k (\gamma +1)_k (2k+ \gamma +\delta +1)},
\end{equation}
see \cite{chich, koek}.

Hence, assuming the frequency resonance condition \eqref{conresonant} for the reduced Hamiltonian \eqref{qH3}, we obtain the explicit expression
\begin{equation}\label{Hdecomposition}
\hat{H} = \sum_{\vec{c}\in C_3} \textbf{R}_{\vec{c}} (\hbar^2 g \textbf{D}_{\vec{c}} + \hbar \lambda_{0\vec{c}} P_{\vec{c}} ) \textbf{R}_{\vec{c}}^T
\end{equation}
for the spectral decomposition of $\hat{H}$ and  thus, the evolution flow
\begin{equation}\label{expqh1}
\exp \left(i\frac{t}{\hbar} \hat{H}\right) = \sum_{\vec{c} \in C_3} \exp \left(it\textbf{H}_{\vec{c}}\right)P_{\vec{c}}, 
\end{equation}
where 
\begin{equation}\label{exphc}
\exp \left(it \textbf{H}_{\vec{c}}\right) = \exp \left(it \lambda_{0\vec{c}}\right) \textbf{R}_{\vec{c}} \exp \left(it\hbar g\textbf{D}_{\vec{c}}\right) \textbf{R}^T_{\vec{c}}
\end{equation}
and $\textbf{D}_{\vec{c}}$ is the diagonal matrix, whose matrix elements are   defined by $\textbf{D}_{kl}^{\vec{c}} := \lambda_k \delta_{kl}$.

In consequence, the solution of Heisenberg equation 
\begin{equation}\label{eqHeis}
\frac{d}{dt}\hat{F}(t) = \frac{i}{\hbar} [\hat{H}, \hat{F}(t)] 
\end{equation}
is also obtained, i.e. 
\begin{equation}\label{evHeis}
\mathbb{R}\ni t \mapsto \hat{F}(t) = e^{-i\frac{t}{\hbar}\hat{H}} F(0) e^{i\frac{t}{\hbar}\hat{H}}\in \mathcal{A}_{\mathcal{G}_{\hbar}}, 
\end{equation}
for some initial condition $\hat{F}(0) \in \mathcal{A}_{\mathcal{G}_{\hbar}}$.

Ending this section, let us mention that the matrix elements of the evolution operator $e^{i\frac{t}{\hbar} \hat{H}}$ with respect to the Fock basis \eqref{basisnew} of $\mathcal{H}$ are the following
\begin{multline}\label{nmtransition}
\langle m, c_1' -n, -c_3'+n , c_2' +c_3' -n |e^{i\frac{t}{\hbar}\hat{H}}|n, c_1 -n, -c_3+n , c_2 +c_3 -n\rangle = \\
\delta_{\vec{c}, \vec{c}'} e^{it\lambda_{0\vec{c}}} \sum_{k=0}^N e^{itg\hbar \lambda_k} R_{nk}^{\vec{c}} R_{mk}^{\vec{c}}.
\end{multline}
The above formula describes explicitly the time dependence of the transition amplitude between Fock states $|n, c_1-n, -c_3 +n, c_2+c_3 -n\rangle $ and $|m, c_1 ' -m , -c_3'+m, c_2'+c_3' -m \rangle $.

Concluding, let us mention that in \cite{3mod} a non-linear three-mode Hamiltonian system was considered and the classical solutions were found, but the spectrum of its quantum Hamiltonian was obtained for particular invariant subspaces only. Opposite to the above case, investigated here four-wave mixing Hamiltonian system is integrated in both versions, classical and quantum.

\section{Correspondence between classical and quantum cases}\label{sec3}

In this section we will show that the quantum four-wave mixing system described by Hamiltonian \eqref{qH1}, in the limit $\hbar \to 0$  goes to the classical one defined by Hamiltonian \eqref{H}. We will also show, that the quantum algebra $\mathcal{A}_{\mathcal{G}_{\hbar}}$ generated by $A_0, A_1, A_2, A_3, A$ and $A^*$ in the limit $\hbar \to 0$ corresponds to the Poisson subalgebra $C^\infty_{\mathcal{G}_0}(\Omega_4)$ defined by the relations (\ref{na1}-\ref{na3}).

For this aim, we recall that the standard coherent states in four-modes case are defined by 
\begin{equation}\label{glauber}
|z_0, z_1, z_2, z_3 \rangle := \sum_{n_0,n_1, n_2, n_3 =0}^\infty \frac{z_0^{n_0} z_1^{n_1} z_2^{n_2} z_3^{n_3}}{\sqrt{n_0! n_1! n_2! n_3!}} \hbar^{-\frac{1}{2} (n_0 +n_1 +n_2 +n_3)} |n_0, n_1, n_2, n_3 \rangle  
\end{equation}
where $(z_0, z_1, z_2, z_3)^T \in \Omega_4$. Let us mention that the map $\mathcal{K} : \Omega_4 \ni (z_0, z_1, z_2, z_3)^T \mapsto \mathbb{C} |z_0, z_1, z_2, z_3\rangle \in \mathbb{C}\mathbb{P} (\mathcal{H})$, called later the standard coherent state map, is a symplectic embedding of $(\Omega_4 , \omega_4)$ into $(\mathbb{C}\mathbb{P} (\mathcal{H}), \omega_{F-S})$, where $\omega_4$ is the canonical symplectic form \eqref{formc4} on $\Omega_4$ and $\omega_{F-S}$ is Fubini-Study form on the complex projective space $\mathbb{C}\mathbb{P}(\mathcal{H})$, e.g. see \cite{CSAO}.

We consider  the operator 
\begin{equation}\label{32}
\hat{F} := \sum_{\overset{m_0, m_1, m_2, m_3,}{n_0, n_1, n_2, n_3 =0}}^\infty f_{m_0, m_1, m_2, m_3, n_0, n_1, n_2, n_3} (a_0^*)^{m_0}(a_1^*)^{m_1}(a_2^*)^{m_2}(a_3^*)^{m_3}a_0^{n_0}a_1^{n_1}a_2^{n_2}a_3^{n_3},
\end{equation}
where $f_{m_0, m_1, m_2, m_3, n_0, n_1, n_2, n_3} \in \mathbb{C}$, expressed by the annihilation and creation operators taken in the normal order. The covariant symbol $\langle \hat{F} \rangle : \mathbb{C}^4 \to \mathbb{C}$  is defined 
\begin{multline}\label{powerser}
\langle \hat{F} \rangle (\bar z_0, \bar z_1, \bar z_2, \bar z_3,z_0, z_1, z_2, z_3)  := \frac{\langle z_0, z_1, z_2, z_3 |F| z_0, z_1, z_2, z_3 \rangle}{ \langle z_0, z_1, z_2, z_3 | z_0, z_1, z_2, z_3 \rangle }=\\
\sum_{\overset{m_0, m_1, m_2, m_3,}{n_0, n_1, n_2, n_3 =0}}^\infty f_{m_0, m_1, m_2, m_3, n_0, n_1, n_2, n_3} ( \bar z_0^*)^{m_0}(\bar z_1^*)^{m_1}(\bar z_2^*)^{m_2}(\bar z_3^*)^{m_3}z_0^{n_0}z_1^{n_1}z_2^{n_2}z_3^{n_3}
\end{multline}
as the mean value function on the coherent states $|z_0, z_1, z_2, z_3\rangle $. In the definition \eqref{32} we have assumed such choice of the coefficients $f_{m_0, m_1, m_2, m_3, n_0, n_1, n_2, n_3}$ which assures the convergence  of the power series \eqref{powerser} on $\mathbb{C}^4$. This means that the coherent states \eqref{glauber} span the domain of $\hat{F}$.  
In the sequel, for simplicity of notation, we will write $f$ and $g$ for the covariant symbols $\langle \hat{F}\rangle$ and $\langle \hat{G} \rangle $.

One defines the $\ast_\hbar$-product of covariant symbols $f,g \in C^\infty (\mathbb{C}^4)$ as 
\begin{equation}\label{starprod}
(f\ast_\hbar g) (\bar z_0, \bar z_1, \bar z_2, \bar z_3,z_0, z_1, z_2, z_3) := \langle \hat{F}\hat{G}\rangle (\bar z_0, \bar z_1, \bar z_2, \bar z_3,z_0, z_1, z_2, z_3) 
\end{equation}
the mean value of the  product $\hat{F}\hat{G}$ of respective operators. 
Substituting  the identity resolution 
\begin{equation}\label{identityres}
\mathbbm{1} = \int_{\mathbb{C}^4} \frac{|z_0, z_1, z_2, z_3 \rangle \langle z_0, z_1, z_2, z_3 |}{\langle z_0, z_1, z_2, z_3 | z_0, z_1, z_2, z_3 \rangle } \frac{1}{(\pi\hbar)^4} dx_0dx_1dx_2dx_3dy_0dy_1dy_2dy_3, 
\end{equation}
where $z_k = x_k + i y_k \in \mathbb{C}$, $k=0,1,2,3$, into \eqref{starprod} one obtains the equivalent formula 
\begin{multline}\label{star2}
(f\ast_\hbar g) (\bar z_0, \bar z_1, \bar z_2, \bar z_3,z_0, z_1, z_2, z_3) =\\
 \sum_{j_0, j_1, j_2, j_3 =0}^\infty \frac{\hbar^{j_0+j_1+j_2+j_3}}{j_0!j_1!j_2!j_3!}\left(\frac{\partial^{j_0}}{\partial z_0^{j_0}}\frac{\partial^{j_1}}{\partial z_1^{j_1}}\frac{\partial^{j_2}}{\partial z_2^{j_2}}\frac{\partial^{j_3}}{\partial z_3^{j_3}}\right) f (\bar z_0, \bar z_1, \bar z_2, \bar z_3,z_0, z_1, z_2, z_3)\times \\
\left(\frac{\partial^{j_0}}{\partial\bar z_0^{j_0}}\frac{\partial^{j_1}}{\partial \bar z_1^{j_1}}\frac{\partial^{j_2}}{\partial \bar z_2^{j_2}}\frac{\partial^{j_3}}{\partial \bar z_3^{j_3}}\right) g (\bar z_0, \bar z_1, \bar z_2, \bar z_3,z_0, z_1, z_2, z_3) 
\end{multline}
for $\ast_\hbar$-product \eqref{starprod}, see \cite{KS}. Let us note here that $\Omega_4$ is an open dense subset of $\mathbb{C}^4$, so, in \eqref{identityres} one can integrate over $\Omega_4$ instead of $\mathbb{C}^4$. From \eqref{star2}, one immediately obtains
\begin{align}
\label{starcor}
f\ast_\hbar g    &\underset{\hbar\to 0}{\rightarrow} f\cdot g, \\
\label{pbcor}
\frac{-i}{\hbar} \left(f\ast_\hbar g   - g\ast_\hbar f\right) & \underset{\hbar\to 0}{\rightarrow} \{f, g\},
\end{align}
where $f \cdot g$ and $\{f, g \}$ are the product and the Poisson bracket \eqref{pb} of the functions $f,g \in C^\infty (\Omega_4)$. 
The correspondences \eqref{starcor} and \eqref{pbcor} allow us to find the one between the subalgebra $\mathcal{A}_{\mathcal{G}_{\hbar}}$ of quantum observables and the Poisson subalgebra $C^\infty_{\mathcal{G}_0}(\Omega_4)$ of classical observables. Namely, from \eqref{clasvar},\eqref{z} and (\ref{qvariables1}-\ref{qvariables5}) we have
\begin{equation}
z =  \langle A \rangle , \quad \bar z =  \langle A^* \rangle \mbox{ and }  I_k=  \langle A_k \rangle, \mbox{ for } k=0,1,2,3,
\end{equation}
for the covariant symbols of $A, A^*$ and $A_k$, $k=0,1,2,3$. Let us note here that $\mathcal{G}_{\hbar} (A_0, A_1, A_2, A_3) \stackrel{\hbar\to 0}{\rightarrow} \mathcal{G}_0 (I_0, I_1, I_2, I_3)$ and $\frac{1}{\hbar}(\mathcal{G}_{\hbar}(A_0 +\hbar , A_1, A_2, A_3) -\mathcal{G}_{\hbar}(A_0  , A_1, A_2, A_3)) \stackrel{\hbar \to 0}{\rightarrow} \frac{\partial \mathcal{G}_0}{\partial I_0} (I_0, I_1, I_2, I_3)$.

Combining the above facts with the relations (\ref{coma0a}-\ref{relcom}), we obtain that if $\hbar \to 0$, then quantum algebra $\mathcal{A}_{\mathcal{G}_{\hbar}}$ corresponds to the Poisson algebra $C^\infty_{\mathcal{G}_0} (\Omega_4)$ generated by classical observables $I_0, I_1, I_2, I_3, z$ and $\bar z$, see (\ref{na1}-\ref{na3}).

In particular,  the classical Hamiltonian \eqref{H}  is obtained as the limit 
\begin{equation}
H= \lim_{\hbar \to 0} \langle \hat{H} \rangle 
\end{equation}
of the coherent state mean value function of the quantum Hamiltonian \eqref{qH1}.

Hence, using \eqref{pbcor} we find that the Heisenberg equation \eqref{eqHeis} and the Heisenberg evolution $\mathbb{R} \ni t \mapsto \hat{F} (t)$ \eqref{evHeis}  in the limit $\hbar \to 0$  correspond  to the Hamilton equation 
\begin{equation}
\frac{d}{dt}f(t) = \{H, f(t)\}
\end{equation}
and  to the Hamiltonian evolution $\mathbb{R}\ni t \mapsto f(t)$ of $f= \lim_{\hbar \to 0} \langle \hat{F} \rangle$.

\section{Reduced coherent state map}\label{sec4}

In this section we combine the classical and quantum reduction procedures in order to construct the reduced coherent state map $\mathcal{K}_{\vec{c}} : \vec{I}^{-1}(\vec{b})/ \mathbb{T}^3 \to \mathbb{C}\mathbb{P} (\mathcal{H}_{\vec{c}})$, where $\mathcal{H}_{\vec{c}} \subset \mathcal{H}$ is Hilbert subspace obtained by the quantum reduction (see Section \ref{sec2}) and  $\vec{I}^{-1}(\vec{b})/ \mathbb{T}^3 \cong ]a,b[\times \mathbb{S}^1$ is the classical reduced phase space (see Section \ref{sec1}).

The standard coherent state map $\mathcal{K}: \Omega_4 \to \mathcal{H}$, see \eqref{glauber}, satisfies the equivariance property 
\begin{equation}\label{flows}
|\sigma_{I_{r}}(t) (z_0, z_1, z_2, z_3 )\rangle = e^{i\frac{t}{\hbar}A_r} |z_0, z_1, z_2, z_3\rangle , 
\end{equation}
where $r=1,2,3$, and $\mathbb{R} \ni t \mapsto \sigma_{I_r}(t) \in \mbox{SpDiff} (\Omega_4, \omega_4)$ and $\mathbb{R}\ni t \mapsto U_{A_r}(t) := e^{i\frac{t}{\hbar} A_r} \in \mbox{Aut} \mathcal{H}$ are classical and quantum flows generated by $I_r$ and $A_r$, respectively.

From the decomposition of $\mathcal{H}$ on the Hilbert subspaces $\mathcal{H}_{\vec{c}}$, where $\vec{c}\in C_3$, follows the decomposition 
\begin{equation}
|z_0, z_1, z_2, z_3 \rangle = \sum_{\vec{c} \in C_3} P_{\vec{c}}|z_0, z_1, z_2, z_3 \rangle 
\end{equation}
of the standard coherent state \eqref{glauber}, where $(z_0, z_1, z_2, z_3)^T\in \Omega_4$. For fixed $\mathcal{H}_{\vec{c}}$ we have 
\begin{equation}\label{projGlauber}
P_{\vec{c}}|z_0, z_1, z_2, z_3 \rangle = \frac{\alpha(z_0, z_1, z_2, z_3)}{\sqrt{\hbar^{2N+\delta +\gamma}}}|\zeta; \vec{c}\rangle, 
\end{equation}
where $\zeta \in \mathbb{C}\backslash \{0\} $ is defined by 
\begin{equation}
\zeta = \frac{z_0z_2}{z_1z_3}
\end{equation}
and $|\zeta; \vec{c}\rangle $  by
\begin{equation}\label{redstate}
|\zeta ; \vec{c}\rangle:= \sum_{n=0}^N \frac{\zeta^n}{\sqrt{n!(N-n)!(n+\gamma)!(N+\delta -n)!}} |n\rangle .
\end{equation}
The complex coefficient $\alpha (z_0, z_1, z_2, z_3) \in \mathbb{C}\backslash \{0\}$ in \eqref{projGlauber}  for subcases (i)-(iv), mentioned in (\ref{cases1}-\ref{cases4}) is given by 
\begin{align}
\label{alfa1}
(i) &\quad \alpha (z_0, z_1, z_2, z_3) = z_1^N z_2^\gamma z_3^{N+\delta}, \\
(ii)& \quad\alpha (z_0, z_1, z_2, z_3) = z_1^{N+\delta} z_2^\gamma z_3^{N},\\
(iii)& \quad\alpha (z_0, z_1, z_2, z_3) = z_0^\gamma z_1^N z_3^{N+\delta},\\
\label{alfa4}
(iv)&  \quad\alpha (z_0, z_1, z_2, z_3) = z_0^\gamma z_1^{N+\delta} z_3^{N}, 
\end{align}
respectively. Restricting the map $\Omega_4 \ni (z_0, z_1, z_2, z_3)^T \mapsto P_{\vec{c}}|z_0, z_1, z_2, z_3 \rangle $ to $\vec{I}^{-1}(\vec{b}) \subset \Omega_4 $, i.e. taking 
\begin{equation}\label{hatz}
\zeta (I_0, \psi_0) = \sqrt{\frac{I_0(I_0-b_3)}{(I_0-b_1)(I_0-b_2-b_3)}} e^{i\psi_0}, 
\end{equation}
and 
\begin{align}
z_0 = &\sqrt{I_0}e^{i(\psi_0 +\psi_1 +\psi_3)},\\
z_1 = &\sqrt{b_1 - I_0} e^{i\psi_1},\\
z_2 = &\sqrt{I_0 - b_3} e^{i(\psi_2 - \psi_3)}, \\
z_3 = &\sqrt{b_2 + b_3 - I_0} e^{i\psi_2},
\end{align}
we obtain the map
\begin{multline}\label{redmap}
]a,b[\times \mathbb{S}^1 \ni (I_0, e^{i\psi_0}) \mapsto \mathcal{K}_{\vec{c}} (I_0, \psi_0 ) := \frac{| \zeta(I_0. \psi_0); \vec{c}\rangle \langle \zeta(I_0. \psi_0); \vec{c}|}{\langle \zeta(I_0. \psi_0); \vec{c}|\zeta(I_0. \psi_0); \vec{c}\rangle}\cong \\
 (\mathbb{C}\backslash \{0\}) |\zeta(I_0. \psi_0); \vec{c} \rangle \in \mathbb{C}\mathbb{P}(\mathcal{H}_{\vec{c}}) , 
\end{multline}
where $\zeta \in \mathbb{C}\backslash \{0\}$ is given by \eqref{hatz}.
This is the reduced  coherent state map $\mathcal{K}_{\vec{c}}:]a,b[\times \mathbb{S}^1 \to \mathbb{C}\mathbb{P} (\mathcal{H}_{\vec{c}})$ of the classical phase space $(]a,b[\times \mathbb{S}^1, \omega )$ into the quantum phase space $(\mathbb{C}\mathbb{P}(\mathcal{H}_{\vec{c}}), \omega_{F-S})$. Note here that the variable $\zeta$ is an invariant of the flows $\sigma_{I_r}(t) $ and, as it follows from \eqref{iflows} and \eqref{flows},  the function $\alpha|_{\vec{I}^{-1}(\vec{b})}$ changes only by the factors $e^{itc_r}$, so, the map \eqref{redmap} is correctly defined on $ \vec{I}^{-1}(\vec{b})/\mathbb{T}^3$.  However, opposite to the standard coherent state map $\mathcal{K}: \Omega_4 \to \mathbb{C}\mathbb{P}(\mathcal{H})$, the reduced coherent state map $\mathcal{K}_{\vec{c}} : ]a,b[ \times \mathbb{S}^1 \to \mathbb{C}\mathbb{P} (\mathcal{H}_{\vec{c}})$ is not a symplectic map, since it only satisfies $\mathcal{K}_{\vec{c}}^*\omega_{F-S} = \rho(I_0) dI_0 \wedge d\psi_0$, where the factor function $\rho ( I_0) $ is not equal to $1$.

As we see from \eqref{redmap}, the presence of the factor $\frac{\alpha (z_0, z_1, z_2, z_3)}{\sqrt{\hbar^{2N+\gamma +\delta }}}$ in \eqref{projGlauber} has no influence on the form of the reduced coherent state map. But, the transition amplitude between Fock state $|n, c_1-n , -c_3+n, c_2+c_3 -n \rangle \in \mathcal{H}_{\vec{c}}$ and a standard coherent state \eqref{glauber} depends on the above factor:
\begin{multline}\label{eq516}
\frac{\langle z_0, z_1, z_2, z_3 |e^{i\frac{t}{\hbar}\hat{H}} | n, c_1-n , -c_3+n, c_2+c_3 -n \rangle }{\sqrt{\langle z_0, z_1, z_2, z_3|z_0, z_1, z_2, z_3\rangle }} =\\
 \frac{\alpha (z_0, z_1, z_2, z_3)}{\hbar^{-\frac{1}{2}(2N+\gamma +\delta)}e^{\frac{1}{2\hbar}(|z_0|^2 +|z_1|^2 +|z_2|^2 +|z_3|^2)}} \langle \zeta ; \vec{c}|e^{it\textbf{H}_{\vec{c}}}|n, c_1-n , -c_3+n, c_2+c_3 -n \rangle=\\
 \frac{\alpha \left(\frac{z_0}{\sqrt{\hbar}}, \frac{z_1}{\sqrt{\hbar}}, \frac{z_2}{\sqrt{\hbar}}, \frac{z_3}{\sqrt{\hbar}}\right)}{e^{\frac{1}{2\hbar}(|z_0|^2 +|z_1|^2 +|z_2|^2 +|z_3|^2)}}e^{it\lambda_{0\vec{c}}} \sum_{k,m=0}^N \frac{e^{itg\hbar \lambda_k}}{\langle \lambda_k |\lambda_k \rangle } \frac{R_n(\lambda_k; \gamma , \delta , N) R_m (\lambda_k ; \gamma , \delta , N) \bar{\zeta}^m }{\sqrt{m! (N-m)! (\gamma +m )! (N + \delta -m )!}}.
\end{multline}
Now, let us notice that after restriction to $\vec{I}^{-1}(\vec{b})$ we have 
\begin{equation}\label{eq517}
\frac{\alpha \left(\frac{z_0}{\sqrt{\hbar}}, \frac{z_1}{\sqrt{\hbar}}, \frac{z_2}{\sqrt{\hbar}}, \frac{z_3}{\sqrt{\hbar}}\right)}{e^{\frac{1}{2\hbar}(|z_0|^2 +|z_1|^2 +|z_2|^2 +|z_3|^2)}} = \frac{\left(\frac{ b_1 - I_0}{\hbar}\right)^{\frac{c_1}{2}}\left(\frac{I_0 - b_3}{\hbar}\right)^{\frac{-c_3}{2}}\left(\frac{b_2+b_3 -I_0}{\hbar}\right)^{\frac{c_2+c_3}{2}}}{e^{\frac{1}{2\hbar}(b_1 +b_2)}}
\end{equation}
in the subcases (i) and (ii) and 
\begin{equation}\label{eq518}
\frac{\alpha \left(\frac{z_0}{\sqrt{\hbar}}, \frac{z_1}{\sqrt{\hbar}}, \frac{z_2}{\sqrt{\hbar}}, \frac{z_3}{\sqrt{\hbar}}\right)}{e^{\frac{1}{2\hbar}(|z_0|^2 +|z_1|^2 +|z_2|^2 +|z_3|^2)}}= \frac{\left(\frac{I_0}{\hbar}\right)^{\frac{c_3}{2}}\left(\frac{b_1 - I_0}{\hbar}\right)^{\frac{c_1-c_3}{2}}\left(\frac{b_2+b_3-I_0}{\hbar}\right)^{\frac{c_2}{2}}}{e^{\frac{1}{2\hbar}(b_1 + b_2)}}
\end{equation}
in the subcases (iii) and (iv). 
One sees from \eqref{eq517} and \eqref{eq518} that if $\frac{1}{\hbar} \vec{b} \to \infty$, i.e. if $\hbar \to 0$, the term $e^{-\frac{1}{2\hbar}(b_1+b_2)}$ dominates in the factor $e^{-\frac{1}{2\hbar}(b_1+b_2)} \alpha|_{\vec{I}^{-1}(\vec{b})}$. Thus and from \eqref{eq516},  we see that the transition amplitude \eqref{eq516} between Fock state and standard coherent state goes to zero if $\hbar \to 0$. Taking into account that Fock states occur only in the context of quantum description and the coherent states are quantum states most similar to the classical ones, i.e. they minimize uncertainty principle, we conclude that the quantum fluctuation of the classical evolution $\mathbb{R} \ni t \mapsto (z_0(t), z_1(t), z_2(t), z_3(t))^T \in \vec{I}^{-1}(\vec{b})$, see section \ref{sec1}, could be neglected in the classical limit $\frac{1}{\hbar} \vec{b} \to \infty $.

We end this section presenting some properties of the reduced coherent state map \eqref{redmap}. Namely, applying the general formula for reproducing measure for the reduced coherent states (proved in Section 5 of \cite{KS}) to the considered here particular case, we obtain the resolution 
\begin{equation}\label{resp}
P_{\vec{c}}  = \int_{\mathbb{C}\backslash \{0\}} \frac{|\zeta; \vec{c}\rangle \langle \zeta; \vec{c}|}{\langle \zeta; \vec{c}|\zeta; \vec{c}\rangle } \langle \zeta; \vec{c}|\zeta; \vec{c}\rangle d\nu_{\vec{c}} (\bar{\zeta}, \zeta) 
\end{equation}
of the orthogonal projection $P_{\vec{c}}$ on the reduced coherent states projections $\frac{|\zeta; \vec{c}\rangle \langle \zeta; \vec{c}|}{\langle \zeta; \vec{c}|\zeta; \vec{c}\rangle }$. The reproducing measure $d\nu_{\vec{c}}$ in \eqref{resp} is given by  
\begin{multline}\label{measure}
d\nu_{\vec{c}} (\bar{\zeta}, \zeta) = \frac{(N+1)! (N+\gamma+1)!(N+\delta+1)!(N+\delta +\gamma +1)!}{2\pi (2N+\delta +\gamma +3)!}\times \\
 \mbox{ }_2F_1\left(\left.\begin{array}{c}
N+\delta +2, N+2 \\
2N+\delta+\gamma +4 , 
\end{array}\right| 1-|\zeta|^2 \right) d|\zeta|^2 d\psi . 
\end{multline}
From \eqref{resp} follows the reproducing property 
\begin{equation}\label{reppropert}
\psi (w) = \int_{\mathbb{C}\backslash \{0\}} \psi(\zeta) \langle \zeta; \vec{c}|\hat{w}; \vec{c}\rangle  d\nu_{\vec{c}} (\bar{\zeta}, \zeta), 
\end{equation}
for $\psi (\hat{w}) := \langle \psi |\hat{w} ; \vec{c}\rangle $, where $|\psi \rangle \in \mathcal{H}_{\vec{c}}$. The reproducing kernel in \eqref{reppropert} is given by
\begin{equation}
\langle \zeta; \vec{c}|\hat{w}; \vec{c}\rangle = \sum_{n=0}^N \frac{(\bar{\zeta} \hat{w})^n}{n!(N-n)!(n+\gamma)!(N+\delta-n)!} = \mbox{ }_2F_1\left(\left.\begin{array}{c}
-N, - (N+\delta) \\
\gamma +1 , 
\end{array}\right| \bar{\zeta} \hat{w} \right) .
\end{equation} 
Note here that the map $|\psi\rangle \mapsto \psi (\zeta) $ defines an antilinear isomorphism between the Hilbert subspace $\mathcal{H}_{\vec{c}}$ and the space $L^2(\mathbb{C}\backslash \{0\} , d\nu_{\vec{c}} )$ of polynomials on $\mathbb{C}\backslash \{0\}$ of degree not greater that $N$ with the scalar product of the $\psi, \phi \in L^2(\mathbb{C}\backslash \{0\} , d\nu_{\vec{c}} )$ defined by 
\begin{equation}
\langle \psi | \phi \rangle = \int_{\mathbb{C}\backslash \{0\}} \overline{\psi(\zeta)} \phi(\zeta) d\nu_{\vec{c}} (\bar{\zeta}, \zeta). 
\end{equation}

From the first equality in \eqref{a0} and \eqref{a}, \eqref{astar}, we obtain the action 
\begin{align}
\textbf{A}_0 |\zeta; \vec{c}\rangle = &\zeta \frac{d}{d\zeta} |\zeta; \vec{c} \rangle, \\
\textbf{A} |\zeta; \vec{c}\rangle = &\zeta\left(N - \zeta \frac{d}{d\zeta}\right)\left(N+\delta - \zeta \frac{d}{d\zeta}\right) |\zeta; \vec{c} \rangle , \\
\textbf{A}^*|\zeta; \vec{c}\rangle = &\frac{d}{d\zeta}\left(\gamma + \zeta \frac{d}{d\zeta}\right) |\zeta; \vec{c} \rangle  
\end{align}
of operators $\textbf{A}_0$,$\textbf{A}$, $\textbf{A}^*$  on the reduced coherent states \eqref{redstate}. 
From the above, one immediately obtains their  action on $\psi \in L^2(\mathbb{C}\backslash \{0\} , d\nu_{\vec{c}} )$: 
\begin{align}
\textbf{A}_0 \psi (\zeta) = \langle \textbf{A}_0\psi | \zeta ; \vec{c} \rangle = &\zeta \frac{d}{d\zeta} \psi(\zeta), \\
\textbf{A} \psi (\zeta) = \langle \textbf{A}\psi | \zeta ; \vec{c} \rangle = & \frac{d}{d\zeta}\left(\gamma + \zeta \frac{d}{d\zeta}\right) \psi( \zeta), \\
\textbf{A}^* \psi (\zeta) = \langle \textbf{A}^*\psi | \zeta ; \vec{c} \rangle = & \zeta\left(N - \zeta \frac{d}{d\zeta}\right)\left(N+\delta - \zeta \frac{d}{d\zeta}\right) \psi ( \zeta) , 
\end{align}
finding in such a way a holomorphic representation of the quantum Kummer shape algebra $\mathcal{A}_{\mathcal{G}_{\hbar}, \vec{c}}$.

\section{Some physical interpretations and applications}\label{sec5}

The natural question concerning possible physical interpretations and applications of the mathematical results obtained in previous sections arises. Since the Hamiltonian system, see \eqref{H} and \eqref{qH1}, is integrated on the classical as well as on the quantum level and the explicit expressions on its mathematical characteristics are obtained, we can expect that this system will be useful also for modeling some physical phenomena.

The first possible physical interpretation of the system, when its Hamiltonian has form \eqref{qH1}, i.e. when it describes the interaction of four radiation modes, is mentioned in the title of the paper.

There are two other possible physical interpretations of the  model given given by Hamiltonian  \eqref{qH1}. In order to discuss them, let us define the following two systems of operators:
\begin{align}
\hat{L}:= &\frac{1}{2} (a_0^*a_0 +a_1^*a_1) ,\\
\hat{M}_+ := &a_0a_1^* , \quad \hat{M}_- := a_0^*a_1 = \hat{M}_+^* , \quad \hat{M}_3 := \frac{1}{2} (a_1^*a_1- a_0^*a_0), \\
\hat{M}_\pm = &\hat{M}_1 \pm i\hat{M}_2
\end{align}
and 
\begin{align}
\hat{R}:= &\frac{1}{2} (a_2^*a_2 +a_3^*a_3) ,\\
\hat{S}_+ := &a_2^*a_3, \quad \hat{S}_- := a_2a_3^* = \hat{S}_+^* , \quad \hat{S}_3 := \frac{1}{2} (a_2^*a_2- a_3^*a_3), \\
\hat{S}_\pm = &\hat{S}_1 \pm i\hat{S}_2 . 
\end{align}
It is easy to see that $\vec{\hat{M}} $ and $\vec{\hat{S}}$ satisfy the commutation relations 
\begin{equation}\label{Mcomrel}
[\hat{M}_k, \hat{M}_l ] =  \varepsilon_{klm} i\hbar \hat{M}_m , \quad [\hat{S}_k, \hat{S}_l ] = \varepsilon_{klm} i\hbar \hat{S}_m, \mbox{ } k,l,m=1,2,3. 
\end{equation}
\begin{equation}
[\hat{M}_k, \hat{S}_l] =0
\end{equation}
for the Lie algebra $so (4, \mathbb{R})\cong so(3, \mathbb{R}) \times so(3, \mathbb{R})$ of the group $SO(4, \mathbb{R}) \cong SO(3, \mathbb{R}) \times SO(3, \mathbb{R})$. Additionally one has
\begin{equation}
[\hat{L}, \hat{S}_k]= [\hat{L}, \hat{M}_k]= [\hat{R}, \hat{S}_k]=[\hat{R}, \hat{M}_k] =[\hat{R}, \hat{L} ]=0 , \mbox{ }k=1,2,3,
\end{equation}
and
\begin{align}
\vec{\hat{M}}^2 = & \hat{L}(\hat{L}+\hbar),\\
\label{eq611}
\vec{\hat{S}}^2 = & \hat{R}(\hat{R}+\hbar).
\end{align}
So,  $\hat{L}$ and $\hat{R}$ as well as $\vec{\hat{M}}^2$ and $\vec{\hat{S}}^2$ are invariants of the Lie group $SO(4, \mathbb{R})$ .

Now, using these new quantum coordinates we rewrite Hamiltonian \eqref{qH1} in the following two ways 
\begin{multline}\label{qHM1}
 \hat{H}_D = (\omega_0 +\omega_1+2\hbar) \hat{L} + (\omega_1 - \omega_0) \hat{M}_3 + \omega_2 a_2^*a_2 + \omega_3 a_3^* a_3 + \\
g [ (a_2^*a_2+a_3^*a_3)\hat{L} + (a_2^*a_2-a_3^*a_3)\hat{M}_3+ a_2a_3^*\hat{M}_+ + a_2^*a_3 \hat{M}_-]
\end{multline}
and 
\begin{multline}\label{qHM2}
 \hat{H}_{MS} = (\omega_0 +\omega_1+ 2\hbar) \hat{L} + (\omega_1 - \omega_0) \hat{M}_3 + (\omega_2 + \omega_3)\hat{R} + (\omega_2 - \omega_3)\hat{S}_3 + \\
g[2\hat{L} \hat{R} + 2\hat{M}_3\hat{S}_3 + \hat{M}_+\hat{S}_- +\hat{M}_-\hat{S}_+]= \\
(\omega_0 +\omega_1+ 2\hbar) \hat{L} + (\omega_1 - \omega_0) \hat{M}_3 + (\omega_2 + \omega_3)\hat{R} + (\omega_2 - \omega_3)\hat{S}_3 + 2g[\hat{L}\hat{R} + \vec{\hat{M}}\cdot \vec{\hat{S}}]. 
\end{multline}

In order to interpret the Hamiltonian \eqref{qHM1}, we recall that in the Dicke model, see e.g. \cite{barry}, where
\begin{equation}
\hat{H}= \omega_2 a_2^* a_2 + \omega_0 \hat{M}_3 + g (a_2^* \hat{M}_- + a_2 \hat{M}_+),
\end{equation}
modeling the system composed of $N$ two-level atoms cooperatively interacting with single photon, the operators $\vec{\hat{M}}$ and $\hat{L}$, satisfying the angular momentum commutation relations \eqref{Mcomrel}, describe the system of atoms while $a_2$ and $a_2^*$ correspond to a photon (wave mode).

Therefore, it is natural to assume that the Hamiltonian \eqref{qHM1} models the interaction of two  radiation quantum modes $a_2$ and $a_3$ with the system of $N$ two-level atoms.

The Hamiltonian \eqref{qHM2} describe a nonlinear interaction of two quantum angular momenta. It is worth to mention in this place that the orthogonal group $SO(4, \mathbb{R}) \cong SO(3, \mathbb{R})\times SO(3, \mathbb{R}) $ could be considered as the dynamical group of the system.

Since $\hat{L}= \frac{1}{2} A_1$, $\hat{R}= \frac{1}{2} A_2$ and $ \hat{M}_3 +\hat{S}_3 = \hat{L}-\hat{R}-A_3$ are integrals of motion for \eqref{qHM2}, so, after assuming the frequency resonance condition \eqref{conresonant}, the interaction described by the Hamiltonian $\hat{H}_{MS}$ reduces to the standard coupling $ \vec{\hat{M}} \cdot \vec{\vec{S}}$ between interacting angular momenta (the other terms are expressed by the integrals of motion). In order to find spectral resolution for $\hat{H}_{MS}$ and evolution flow $\mathbb{R} \ni t \mapsto e^{i\frac{t}{\hbar}\hat{H}_{MS}}$, it is enough to rewrite the Fock base \eqref{newbasis} in terms of the common eigenvectors of the operators $\hat{L}, \hat{M}_3, \hat{R}$ and $\hat{S}_3$.  The other question, important for some physical applications, concerns the relation between the atomic coherent states \cite{arechi} (naturally related to the reduced Hamiltonian $\textbf{H}_{MS, \vec{c}}$) and the reduced coherent states \eqref{redstate} described in Section \ref{sec4}. All above remarks about the Hamiltonian $\hat{H}_{MS}$ after some modifications one can repeat for the Hamiltonian $\hat{H}_D$ also.

Taking the covariant symbols $\langle \hat{H}_D \rangle $ and $\langle \hat{H}_{MS} \rangle $ of the Hamiltonians \eqref{qHM1} and \eqref{qHM2}, in the limit $\hbar \to 0$ we obtain their classical counterparts
\begin{align}
\nonumber
H_D =& (\omega_0 + \omega_1) \mathcal{L} + (\omega_1 - \omega_0) \mathcal{M}_3 + \omega_2 |z_2|^2 + \omega_3 |z_3|^2 + \\
 &g [ (|z_2|^2 + |z_3|^2 )\mathcal{L} + (|z_2|^2 - |z_3|^3 )\mathcal{M}_3 + z_2 \bar z_3 \mathcal{M}_+ + \bar z_2 z_3 \mathcal{M}_-], \\
\nonumber
H_{MS} = &(\omega_0 + \omega_1) \mathcal{L} + (\omega_1 - \omega_0) \mathcal{M}_3 + (\omega_2 +\omega_3 )\mathcal{R} + (\omega_2 - \omega_3)\mathcal{S}_3 + \\
 & 2g[ \mathcal{L}\mathcal{R} + \vec{\mathcal{M}} \cdot \vec{\mathcal{S}}]
\end{align}
where the functions $\mathcal{L}, \mathcal{M}_3, \mathcal{M}_\pm , \mathcal{R}, \mathcal{S}_3, \mathcal{S}_\pm \in C^\infty (\Omega_4) $ defined by 
\begin{equation}
\begin{array}{rl}
\mathcal{L}& := \frac{1}{2} (|z_0|^2 + |z_1|^2 ),\\
\mathcal{M}_3 & := \frac{1}{2} (|z_1|^2 - |z_0|^2 ),\\
\mathcal{M}_+ &:= z_0\bar z_1 , \quad \mathcal{M}_- := \bar z_0 z_1,\\ 
\mathcal{M}_\pm &:=  \mathcal{M}_1 \pm i\mathcal{M}_2,
\end{array}
\mbox{ and } \begin{array}{rl}
\mathcal{R}& := \frac{1}{2} (|z_2|^2 + |z_3|^2 ),\\
\mathcal{S}_3 & := \frac{1}{2} (|z_2|^2 - |z_3|^2 ),\\
\mathcal{S}_+ &:= \bar z_2 z_3 , \quad \mathcal{S}_- := z_2 \bar z_3,\\ 
\mathcal{S}_\pm &:=  \mathcal{M}_1 \pm i\mathcal{M}_2
\end{array}
\end{equation}
satisfy the relations   
\begin{equation}
\{\mathcal{M}_k, \mathcal{M}_l\} = \varepsilon_{klm} \mathcal{M}_m , \quad \{\mathcal{S}_k, \mathcal{S}_l\} = \varepsilon_{klm} \mathcal{S}_m,
\end{equation}
\begin{equation}
\{\mathcal{M}_k, \mathcal{L}\}=\{\mathcal{S}_k, \mathcal{R}\}=\{\mathcal{M}_k, \mathcal{R}\}=\{\mathcal{S}_k, \mathcal{L}\}=\{\mathcal{L}, \mathcal{R}\}=0,
\end{equation}
\begin{equation}
\vec{\mathcal{M}}^2 = \mathcal{L}^2, \quad \vec{\mathcal{S}}^2 = \mathcal{R}^2,
\end{equation}
 which are the classical counterparts of the ones presented in (\ref{Mcomrel}-\ref{eq611}).

The discussion of the physical aspects concerning these Hamiltonians we leave for a next paper. Now however, let us present a few formulas for time evolution of transition probabilities between some Fock states. More precisely,  from \eqref{nmtransition}, one immediately obtains the following general expression 
\begin{multline}
|\langle m, c_1' - m, -c_3' +m , c_2' + c_3 ' -m|e^{it\textbf{H}_{\vec{c}}}|n, c_1 - n, -c_3 +n , c_2 + c_3  -n\rangle|^2 =\\
\delta_{\vec{c}, \vec{c}'}\sum_{k=0, l\leq k}^N 2\cos (g\hbar \lambda_k t) R_{nk}^{\vec{c}} R_{mk}^{\vec{c}}R_{nl}^{\vec{c}}R_{ml}^{\vec{c}}
\end{multline}
for time dependence of transition probability between Fock states $|m, c_1' - m, -c_3' +m , c_2' + c_3 ' -m \rangle$ and $|n, c_1 - n, -c_3 +n , c_2 + c_3  -n \rangle$.

In the case $N=1$,i.e. when Hilbert subspace $\mathcal{H}_{\vec{c}}$ is spanned by two vectors 
\begin{equation}\label{basisex}
|0\rangle = |0, 1, \gamma , \delta +1 \rangle , \quad |1\rangle = |1, 0, \gamma +1 , \delta \rangle ,
\end{equation}
where $\gamma \geq 0 $ and $ \delta \geq 0$, the eigenvalues $\tilde{\lambda}_0$ and $\tilde{\lambda}_1$ of the reduced Hamiltonian $\textbf{H}_{\vec{c}}$ are given by 
\begin{equation}
\tilde{\lambda}_0 = \lambda_{0\vec{c}} , \quad \tilde{\lambda}_1 = g\hbar (\gamma +\delta +2) + \lambda_{0\vec{c}}, 
\end{equation}
where, depending from the subcases (\ref{1case}-\ref{4case}), one has 
\begin{align}
\mbox{(i) } \lambda_{0\vec{c}} = & \omega_1+\omega_3 + \omega_2 \gamma + \omega_3 \delta , \\
\mbox{(ii) } \lambda_{0\vec{c}} = & \omega_1+\omega_3 + \omega_2 \gamma + \omega_1 \delta +g\hbar \delta (\gamma +1), \\
\mbox{(iii) } \lambda_{0\vec{c}} = & \omega_1+\omega_3 + (\omega_1 -\omega_2 +\omega_3)\gamma + \omega_3 \delta +g\hbar \gamma (\delta +1), \\
\mbox{(iv) } \lambda_{0\vec{c}} = & \omega_1+\omega_3 + (\omega_1 -\omega_2 +\omega_3) \gamma + \omega_1 \delta +g\hbar (\gamma +\delta ). 
\end{align}
Therefore operator $\exp (it\textbf{H}_{\vec{c}})$ written in the basis \eqref{basisex} is given by the following $2\times 2$ matrix 
\begin{equation}
e^{it\textbf{H}_{\vec{c}}} = \frac{e^{it\lambda_{0\vec{c}}}}{\lambda_1} \left(\begin{array}{cc}
(\delta +1) + (\gamma +1)e^{i\hbar g \lambda_1 t} & \sqrt{(\gamma+1)(\delta +1)}(1-e^{i\hbar g \lambda_1 t})\\
\sqrt{(\gamma+1)(\delta +1)}(1-e^{i\hbar g \lambda_1 t}) & (\gamma +1)+(\delta +1)e^{i\hbar g \lambda_1 t}
\end{array}\right),
\end{equation}
where $\lambda_1 =\gamma + \delta +2$. Therefore, we are able to find the following transition probabilities between Fock states:\\
(i) the probability the system prepared in a state $|0\rangle $ (or $|1\rangle $) after time $t$ will does not leave this state 
\begin{equation}\label{extp1}
|\langle 0|e^{it\textbf{H}_{\vec{c}}}|0\rangle|^2 = |\langle 1|e^{it\textbf{H}_{\vec{c}}}|1\rangle|^2 =\frac{1+\left(\frac{\gamma +1}{\delta +1 }\right)^2}{\left(1+\frac{\gamma +1}{\delta +1}\right)^2} + \frac{2\frac{\gamma +1}{\delta +1 }}{\left(1+\frac{\gamma +1}{\delta +1}\right)^2} \cos (g\hbar \lambda_1 t),%\frac{1}{\lambda_1^2}((\delta+1)^2 + (\gamma +1)^2 + 2(\delta +1)(\gamma +1) \cos (\hbar g \lambda_1 t)),
\end{equation}
(ii) the probability that the system prepared in state $|0 \rangle $( or $|1\rangle $) after time $t$ will pass to the state $|1\rangle $ ( or $|0\rangle $)
\begin{equation}\label{eqjakies}
|\langle 1|e^{it\textbf{H}_{\vec{c}}}|0\rangle|^2 = |\langle 0|e^{it\textbf{H}_{\vec{c}}}|1\rangle|^2 =\frac{2\frac{\gamma +1}{\delta +1 }}{\left(1+\frac{\gamma +1}{\delta +1}\right)^2}  (1- \cos (g\hbar \lambda_1 t)). %2\frac{(\gamma +1)(\delta +1)}{\lambda_1^2}(1- \cos (\hbar g \lambda_1t )).
\end{equation}
Let us note here that \eqref{eqjakies} describes the time dependence of probability of  conversion  
\begin{equation}
|1, 0, \gamma +1, \delta \rangle \mapsto |0, 1, \gamma , \delta +1 \rangle 
\end{equation}
of two photons of frequencies $\omega_0$ and $\omega_2$ into two photons of frequencies $\omega_1$ and $\omega_3$
or  the opposite conversion
\begin{equation}
|0, 1, \gamma , \delta +1 \rangle  \mapsto |1, 0, \gamma +1, \delta \rangle . 
\end{equation}
of these photons.

As we see the transition probabilities \eqref{extp1} and \eqref{eqjakies} oscillate  with frequency $\hbar g (\gamma + \delta +2)$ and the amplitude of the oscillations is some rational function of parameters  $\gamma \geq 0$ and $\delta \geq 0$. Moreover, amplitude of the oscillations is invariant with respect to the replacement $(\gamma , \delta ) \mapsto (\delta , \gamma )$. Note also, that if we fix one of these parameters, then probability transition \eqref{eqjakies} tends to zero as the second parameter rises.

\end{document}